\documentclass{svmult}
\usepackage{makeidx}         
\usepackage[bottom]{footmisc}

\makeindex             

\usepackage{supertech-lncs}
\usepackage{wrapfig, floatflt}

\newcommand{\mypara}[1]{{\smallskip\noindent {\bf #1}}\quad}

\input{textlater}

\newenvironment{short-enum}{\begin{list}{\thestp .}{\usecounter{stp}
                                \setlength{\itemsep}{0in}
                                \setlength{\topsep}{0in}
                                \setlength{\parsep}{0in}
                                \addtolength{\listparindent}{-2ex}
                                \setlength{\itemindent}{\listparindent}}}%
{\end{list}}

\newenvironment{short-item}{\begin{list}{$\bullet$}{%
                                \setlength{\itemsep}{0in}
                                \setlength{\topsep}{0.03in}
                                \setlength{\parsep}{0in}
                                \addtolength{\listparindent}{-2ex}
                                \setlength{\itemindent}{\listparindent}}}%
{\end{list}}

\newenvironment{flush-item}{\begin{list}{$\bullet$}{%
                                \setlength{\itemsep}{0in}
                                \setlength{\topsep}{0.03in}
                                \setlength{\leftmargin}{0.6\leftmargin}}}%
{\end{list}}

\newenvironment{second-flush-item}{\begin{list}{-}{%
                                \setlength{\itemsep}{0in}
                                \setlength{\topsep}{0.03in}
                                \setlength{\leftmargin}{0.6\leftmargin}}}%
{\end{list}}

\newenvironment{flush-enum}{\begin{list}{\thestp .}{\usecounter{stp}
                                \setlength{\itemsep}{0in}
                                \setlength{\topsep}{0.03in}
                \setlength{\parsep}{1.0\parsep}
                \setlength{\itemindent}{0.8em}
                                \setlength{\leftmargin}{0.6\leftmargin}}}%
{\end{list}}

\newenvironment{short-flush-item}{\begin{list}{$\bullet$}{%
                                \setlength{\itemsep}{0in}
                                \setlength{\topsep}{0.03in}
                                \setlength{\parsep}{0in}
                                \setlength{\leftmargin}{0.6\leftmargin}
                                \addtolength{\listparindent}{-2ex}
                                \setlength{\itemindent}{\listparindent}}}%
{\end{list}}

\newcounter{algline}

\newcommand{\nl}{\\ \thealgline \stepcounter{algline}}

\newcounter{stp}

\newenvironment{boxfig}[2]{\begin{figure*}[#1]\fbox{\begin{minipage}{\linewidth}
                        \vspace{0em}
                        \makebox[0.025\linewidth]{}
                        \begin{minipage}{0.95\linewidth}
                        #2
                        \end{minipage}
                        \vspace{1em}
                        \end{minipage}}}{\end{figure*}}

\newcommand{\vs}[1]{\mbox{\textit{V}}({#1})}
\newcommand{\Voronoi}[1]{\mbox{\textrm{Voronoi}}({#1})}
\newcommand{\vor}[1]{\Voronoi{#1}}
\newcommand{\snow}{s}  
\newcommand{\distance}{d}  
\newcommand{\dist}{\distance}
\newcommand{\cost}{c} 
\newcommand{\size}{s} 

\newcommand{\vdist}[1]{\delta({#1})}

\def\hyperspc{\kern -0.22em}

\newcommand{\remove}[1]{}

\newsavebox{\sq}
\sbox{\sq}{\begin{picture}(10,10)
\put(0,0){\line(1,0){10}}\put(0,0){\line(0,1){10}}
\put(0,10){\line(1,0){10}}\put(10,0){\line(0,1){10}}
\end{picture}}

\newcommand{\mycaption}[1]{\caption{\small{#1}}} 
\newcommand{\old}[1]{}
\newcommand{\novor}[1]{} 

\newcommand{\df}{\left\lfloor\frac{D}{2}\right\rfloor}

\title*{The Snowblower Problem}

\author{Esther M. Arkin\inst{1}\and
Michael A. Bender\inst{2}\and Joseph S. B. Mitchell\inst{1}\and
Valentin Polishchuk\inst{1}}\authorrunning{Esther M. Arkin, Michael
A. Bender, Joseph S. B. Mitchell, Valentin Polishchuk}

\institute{Department of Applied Mathematics and Statistics, Stony
Brook University, Stony Brook, NY 11794, USA.
\texttt{\{estie,jsbm,kotya\}@ams.stonybrook.edu}.
\footnote{Partially supported by the U.S.-Israel Binational Science
Foundation (2000160), NASA (NAG2-1620), NSF (CCR-0098172,
ACI-0328930, CCF-0431030), and Metron Aviation.} \and Department of
Computer Science, Stony Brook University, Stony Brook, NY 11794,
USA. \texttt{bender@cs.stonybrook.edu} \footnote{Partially supported
by NSF Grants EIA-0112849 and CCR-0208670.} }

\begin{document}

\maketitle

\thispagestyle{empty}

\begin{abstract}
We introduce the \emph{snowblower problem} (\emph{SBP}), a new
optimization problem that is closely related to milling problems and
to some material-handling problems. The objective in the SBP is to
compute a short tour for the snowblower to follow to remove all the
snow from a domain (driveway, sidewalk, etc.).  When a snowblower
passes over each region along the tour, it displaces snow into a
nearby region. The constraint is that if the snow is piled too high,
then the snowblower cannot clear the pile.

We give an algorithmic study of the SBP. We show that in general, the
problem is NP-complete, and we present polynomial-time approximation
algorithms for removing snow under various assumptions about the
operation of the snowblower. Most commercially-available snowblowers
allow the user to control the direction in which the snow is thrown.
We differentiate between the cases in which the snow can be thrown in
any direction, in any direction except backwards, and only to the
right. For all cases, we give constant-factor approximation
algorithms; the constants increase as the throw direction becomes more
restricted.

Our results are also applicable to robotic vacuuming (or lawnmowing)
with bounded capacity dust bin and to some versions of
material-handling problems, in which the goal is to rearrange cartons
on the floor of a warehouse.

\end{abstract}


\section{Introduction}

During a recent major snowstorm in the northeastern USA, one of the
authors used a snowblower to clear an expansive driveway.  A
snowblower is a ``material shifting machine'', lifting material from
one place, and depositing it nearby.  There is a skill in making sure
that the deposited piles of snow do not grow higher than the maximum
depth capacity of the snowblower.  This experience crystallized into
an algorithmic question we have called the \emph{Snowblower Problem}
(\emph{SBP}):
\textbf{How does one optimally use a snowblower to clear a given
polygonal region?}

The SBP shows up also in other contexts: Consider a mobile robot that
is equipped with a device that allows it to pick up a carton and then
place the carton down again, possibly on a stack of cartons, in a
location just next to it; with each such operation, the robot shifts a
unit of ``material''.  The SBP models the problem in which the robot
is to move a set of boxes to a specified destination in the most
efficient manner, subject to the constraint that it cannot stack boxes
higher than a capacity bound.

In a third motivating application, consider a robotic lawnmower or
vacuum cleaner that has a catch basin for the clippings, leaves, dust,
or other debris.  The goal is to remove the debris from a region, with
the constraint that the catch basin must be emptied (e.g., in the
compost pile) whenever it gets full.

The SBP is related to other problems on milling, vehicle routing, and
traveling salesman tours, but there are two important new features:
(a) material must be moved (snow must be thrown), and (b) material may
not pile up too high.

While the SBP arises naturally in these other application domains, for
the rest of the paper, we use the terminology of snow removal.

The objective of the SBP is to find the shortest snowblower tour that
clears a domain $P$, assumed to be initially covered with snow at
uniform depth 1.  An important parameter of the problem is the maximum
snow depth $D>1$ through which the snowblower can move.  At all times
no point of $P$ should have snow of greater depth than $D$.  The snow
is to be moved to points outside of $P$. We assume that each point
outside $P$ is able to receive arbitrarily much snow (i.e., that the
driveway is surrounded by a ``cliff'' over which we
can toss as much snow as we want).\footnote{The ``cliff'' assumption
accurately models the capacitated vacuum cleaner problem for which
there is a (central) ``dustpan vac'' in the baseboard, where a robotic vacuum
cleaner may empty its load~\cite{dustpan-inlet} and applies also
to urban snow removal using snow melters~\cite{snow-melter}.}  

Snowblowers offer the user the ability to control the direction in
which the snow is thrown. Some throw directions are preferable over
others; e.g., throwing the snow back into the user's face is
undesirable.  However, it can be cumbersome to change the throw
direction too frequently during the course of clearing; also, some
variants (e.g., lawnmowing with clippings) require that the throw
direction is fixed.  Thus, we consider three {\em throw models}. In
the \emph{default} model throwing the snow backwards is allowed.  In
the \emph{adjustable-throw} model the snow can be thrown only to the
left, right, or forward.  In the \emph{fixed-throw} model the snow is
always thrown to the right.  Even though it seems silly to allow the
throw direction to be back into one's face, the default model is the
starting point for the analysis of other models and is equivalent to
the vacuum cleaner problem (discussed later).

\mypara{Results.} In this paper we introduce the snowblower problem,
model its variants, and give the first algorithmic results for its
solution. We show that the SBP is NP-complete for multiply connected
domains~$P$.  Our main results are constant-factor approximation
algorithms for each of the three throw models, assuming $D\geq 2$;
refer to \tabref{results-back-throw}.  The approximation ratio of our
algorithms increases as the throw direction becomes more restricted.
We give extensions for clearing polygons with holes, both where the
holes are obstacles and cliffs.  Then we discuss how to adapt our
algorithms for clearing nonrectilinear polygons and polygons with
uneven initial distributions of snow.  We conclude by giving a
succinct representation of the snowblower tour, in which the tour
specification is polynomial in the complexity of the input polygon.

\begin{table}[hbt]
\centering
 {\footnotesize
\begin{tabular}{|l|c|c|} \hline
$D$&2 or 3&any~$D\ge4$\\
\hline%
Apx. &6&8\\
\hline%
Ref.&\multicolumn{2}{|c|}{Theorem~\ref{thm_apx_back_throw}}\\
\hline
\end{tabular}}~~~~~%
 {\footnotesize
 \begin{tabular}{|l|c|} \hline
$D$&any~$D\ge2$\\
\hline%
Apx. & $4+{3D}/{\lfloor{D}/{2}\rfloor}$\\
\hline%
Ref.&{Theorem~\ref{thm_apx_adj}}\\
\hline
\end{tabular}}~~~~~%
 {\footnotesize
\begin{tabular}{|l|c|}
\hline
$D$& any $D\ge 2$\\
\hline%
Apx. & $34+24D/\lfloor{D}/{2}\rfloor$\\
\hline%
Ref.& Theorem~\ref{thm_apx_fixed}\\
\hline
\end{tabular}
 } \mycaption{Approximation factors of our algorithms in the default
model (left), adjustable-throw model (middle), and fixed-throw
model (right).}\tablabel{results-back-throw}
\end{table}

\mypara{Related Work.}%
The SBP is closely related to milling and lawn-mowing problems, which
have been studied extensively in the NC-machining and
computational-geometry literatures; see
e.g.,~\cite{ahs-oprzp-00,afm-aalmm-00,h-cgpm-91}. The SBP is also
closely related to material-handling problems, in which the goal is to
rearrange a set of objects (e.g., cartons) within a storage facility;
see~\cite{PushingBlocks_CGTA,c-spc-99,tbmp-p-04}.  The SBP may be
considered as an intermediate point between the
TSP/\allowbreak{lawnmowing}/\allowbreak{milling} problems and
material-handling problems. Indeed, for $D=\infty$, the SBP is that of
optimal milling. Unlike most material-handling problems, the SBP
formulation allows the material (snow) to pile up on a single pixel,
and it is this compressibility of the material that distinguishes the
SBP from previously studied material-handling problems. With TSP and
related problems, a pixel only is visited a constant number of times,
whereas with material-handling problems, pixels may have to be visited
a number of times exponential in the input size. For this reason,
material-handling problems are not even known to be in
NP~\cite{PushingBlocks_CGTA,c-spc-99}, in contrast with the SBP.  Note
that in material handling problems the objective is to minimize {\em
  workload} (distance traveled while loaded), while in the SBP (as in
the milling/mowing problems) the objective is to minimize total travel
distance (loaded or not).

The SBP is also related to the earth-mover's distance (EMD), which is
the minimum amount of work needed to rearrange one distribution (of
earth, snow, etc.) to another; see \cite{cg-emdlbiut-97}.  In the EMD
literature, the question is explored mostly from an existential point
of view, rather than planning the actual process of rearrangement.  In
the SBP, we are interested in optimizing the length of the tour, and
we do not necessarily know in advance the final distribution of the
snow after it has been removed from~$P$.

\mypara{Notation.} The input is a polygonal domain, $P$. Since we are
mainly concerned with proving constant factor approximation
algorithms, it suffices to consider distances measured according to
the $L_1$ metric. We consider the snowblower to be an (axis-parallel)
unit square that moves horizontally or vertically by unit steps.  This
justifies our assumption, in most of our discussion, that $P$ is an
integral-orthogonal simple polygon, which is comprised of a union of
\emph{pixels} -- (closed) unit squares with disjoint interiors and
integral coordinates.  In Section~\ref{sec_extensions} we remark how
our methods extend to general (nonrectilinear) regions.

We say that two pixels are \emph{adjacent} or \emph{neighbors} if they
share a side; the \emph{degree} of a pixel is the number of its
neighbors.
For a region $R\subseteq P$ (subset of pixels), let $G_R$ denote the
\emph{dual graph} of $R$, having a vertex in the center of each pixel
of $R$ and edges between adjacent pixels.  Sometimes when we
speak of the region $R$, we implicitly mean the dual graph $G_R$.
A pixel of degree less than four is a \emph{boundary pixel}. For a
boundary pixel, a side that is also on the boundary of $P$ is called a
\emph{boundary side}.  The set of boundary sides, $\partial P$, forms
the boundary of $P$.  We treat $\partial P$ as a set of
boundary sides, rather than just as a closed curve. We assume that
the elements of $\partial P$ are ordered as they are encountered when
the boundary of $P$ is traversed counterclockwise.

An \emph{articulation vertex} of a graph $G$ is a vertex whose removal
disconnects $G$. We assume that $G_P$ has no articulation vertices.
(Our algorithms can be adapted to regions having articulation
vertices, at a possible increase in approximation ratio.)

\mypara{Algorithms Overview.}%
Our algorithms proceed by clearing the polygon
Voronoi-cell-by-Voronoi-cell, starting from the Voronoi cell of the
garage $g$ --- the pixel on the boundary of $P$ at which the
snowblower tour starts and ends. The order of the boundary sides in
$\partial P$ provides a natural order in which to clear the cells.
We observe that the Voronoi cell of each boundary side is a tree of
one of two special types, which we call \emph{lines} and
\emph{combs}. We show how to clear the trees efficiently in each of
the throw models. We prove that our algorithms give constant-factor
approximations by charging the lengths of the tours produced by the
algorithms to two lower bounds, described in the next section.

\section{Preliminaries}

\mypara{Tie Breaking.}%
Our rules for finding $\vs p$ for a pixel $p$ that is equidistant
between two or more boundaries is based on the direction of the
shortest path from $p$ to $\vs p$; horizontal edges are preferred to
vertical, going down has higher priority than going up, going to the
right --- than going left.

In details, let $p$ be a pixel, equidistant from $m$ boundary sides
$\{e_1\ldots e_m\}=\cal E\subset\partial P$. Let $r(p), l(p), d(p),
u(p)$ denote the pixels to the left, above, to the right and below
$p$.  First consider the case when $p$ is a boundary pixel. If
$r(p)\notin P$, then the right side of $p$ is defined to be $\vs p$;
else if $l(p)\notin P$, the left side of $p$ is $\vs p$; else if
$d(p)\notin P$, the bottom side of $p$ is $\vs p$; otherwise, the
top side of $p$ is $\vs p$.  If $p$ is not a boundary pixel, we
build the path from $p$ to $\vs p$ as follows. If there exists a
side $e\in\cal{E}$ such that there exist a (shortest) path from $p$
to $e$, going through $r(p)$ then $\vs p=\vs{r(p)}$; else if there
exists a side $e\in\cal{E}$ such that there exist a (shortest) path
from $p$ to $e$, going through $l(p)$ then $\vs p=\vs{l(p)}$; else
if there exists a side $e\in\cal{E}$ such that there exist a
(shortest) path from $p$ to $e$, going through $d(p)$ then $\vs
p=\vs{d(p)}$; otherwise $\vs p=\vs{u(p)}$.

We first describe the rules for finding the closest boundary side
for pixels that are equidistant between two or more boundaries.  We
give preference to the boundary sides as follows: horizontal sides
are preferred over vertical.  The lowest horizontal side has the
highest priority among horizontal sides, and among all lowest
horizontal sides, the leftmost has the highest priority.  The
leftmost vertical side has the highest priority among vertical
sides, and among all leftmost vertical sides, the highest has the
highest priority. In fact, any tie-breaking rule can be applied as
long as it is applied consistently.  The particular choice of the
rule only affects the orientation of the combs.
\figlater{
\begin{figure}[h!]
\begin{picture}(200,200)
\put(0,60){\fbox{\large{Tie}}}%
\put(40,90){\fbox{Hor. edge}}\put(40,30){\fbox{Vert. edge}}
\put(110,110){\fbox{down}}\put(110,80){\fbox{up\quad}}
\put(110,50){\fbox{left\,\,\,}}\put(110,20){\fbox{right}}
\put(170,150){\fbox{\small{left\,\,\,}}}\put(170,130){\fbox{\small{right}}}
\put(170,100){\fbox{\small{left\,\,\,}}}\put(170,80){\fbox{\small{right}}}
\put(170,60){\fbox{\small{up\,\,\,}}}\put(170,40){\fbox{\small{down}}}
\put(170,20){\fbox{\small{up\,\,\,}}}\put(170,0){\fbox{\small{down}}}
\put(25,75){\line(1,1){10}}\put(25,55){\line(1,-1){10}}
\put(93,102){\line(2,1){13}}\put(93,85){\line(1,0){13}}
\put(95,40){\line(2,1){11}}\put(95,25){\line(1,0){11}}
\put(142,122){\line(1,1){25}}\put(142,107){\line(1,1){25}}
\put(142,85){\line(6,5){25}}\put(142,80){\line(1,0){25}}
\put(142,55){\line(5,1){25}}\put(142,50){\line(4,-1){25}}
\put(142,25){\line(1,0){25}}\put(142,20){\line(2,-1){25}}
\end{picture}
\begin{picture}(100,100)(-30,-50)
\multiput(0,20)(0,10){3}{\multiput(0,0)(10,0){7}{\usebox\sq}}
\multiput(20,0)(10,0){3}{\multiput(0,0)(0,10){7}{\usebox\sq}}
\put(32,33){$\scriptsize{p}$}\put(12,12){$e$}
\linethickness{2pt}\put(10,20){\line(1,0){10}}\thinlines
\end{picture}
\centering \mycaption{Left: the tie breaking tree; on a tie, the
upper edge of the tree is followed.  Right: an example; by the tie
breaking rules, $\vs p=e$, the bold edge;
$\vdist{p}=4$.}\figlabel{tie_breaking}
\end{figure}
} 

\mypara{Voronoi Cell Structure.}
For a pixel $p\in P$ let $\vs{p}$ denote the element of $\partial
P$, closest to $p$.  In case of ties, the tie-breaking rule is
applied. Inspired by computational-geometry terminology, we call
$\vs{p}$ the \emph{Voronoi side} of $p$.  We let $\vdist{p}$ denote
the length of the path from $p$ to the pixel having $\vs{p}$ as a
side (see \figref{tie_breaking}, right). For a boundary side
$e\in\partial P$ we let $\Voronoi{e}$ denote the (possibly, empty)
set of pixels, having $e$ is the Voronoi side:
$\Voronoi{e}=\set{p\in P|\vs{p}=e}$. We call $\Voronoi e$ the
\emph{Voronoi cell} of $e$. The Voronoi cells of the elements of
$\partial P$ form a partition of $P$, which is called the
\emph{Voronoi decomposition} of $P$.

A set of pixels $\cal L$ whose dual graph $G_{\cal L}$ is a straight
path or a path with one bend, is called a {\em line}.
Each line $\mathcal{L}$ has a {\em root} pixel $p$, which
corresponds to one of the two leaves of $G_\mathcal{L}$, and a
{\em base}, $e\in\partial P$, which is a side of $p$.

A {\em (horizontal) comb} $\mathcal{C}$ is a union of pixels
consisting of a set of vertically adjacent (horizontal) rows of
pixels, with all of the rightmost pixels (or all of the leftmost
pixels) in a common column.  (A vertical comb is defined similarly;
however, by our tie breaking rules, we need consider only horizontal
combs.)  A comb is a special type of {\em histogram}
polygon~\cite{csw-fmasp-99}.  The common vertical column of
rightmost/leftmost pixels is called the {\em handle} of comb
$\mathcal{C}$, and each of the rows is called a {\em tooth}.  A {\em
  leftward} comb has its teeth extending leftwards from the handle; a
{\em rightward} comb is defined similarly.  The pixel of a tooth that
is furthest from the handle is the {\em tip} of the tooth.  The
topmost row is the {\em wisdom tooth} of the comb.  The {\em root}
pixel $p$ of the comb is either the bottommost or topmost pixel of the
handle, and its bottom or top side, $e\in\partial P$, is the {\em
  base} of the comb.  See \figref{comb_and_vor}, left.  The union of a leftward comb
and a rightward comb having a common root pixel is called a {\em
  double-sided comb}.

\begin{figure}
\centering
\begin{picture}(100,100)
\multiput(50,10)(10,0){5}{\multiput(0,0)(0,10){7}{\usebox{\sq}}}
\multiput(10,10)(10,0){4}{\multiput(0,0)(0,10){2}{\usebox{\sq}}}
\multiput(0,30)(10,0){5}{\multiput(0,10)(0,10){2}{\usebox{\sq}}}
\linethickness{2pt}\put(90,10){\line(1,0){10}}\thinlines%
\put(105,40){$c$}
\multiput(93,13)(0,10){7}{$\ast$}\multiput(83,72)(-10,0){4}{$\bullet$}
\end{picture}
\hspace{2.5cm}
\begin{picture}(100,80)(-20,-20)
\multiput(-10,20)(0,10){4}{\multiput(0,0)(10,0){7}{\usebox\sq}}
\multiput(20,-10)(10,0){4}{\multiput(0,0)(0,10){4}{\usebox\sq}}
\put(-7,12){1}%
\put(-7,22){\scriptsize{1}}%
\multiput(3,22)(0,10){2}{\scriptsize{2}}
\multiput(13,22)(10,0){3}{\scriptsize{3}}
\multiput(13,32)(10,0){2}{\scriptsize{3}}
\multiput(23,12)(10,0){2}{\scriptsize{4}}%
\put(23,2){\scriptsize{5}} \put(23,-8){\scriptsize{7}}
\multiput(33,-8)(0,10){2}{\scriptsize{8}}
\multiput(43,-8)(0,10){2}{\scriptsize{9}}
\put(52,-8){\scriptsize{10}}
\end{picture}
\mycaption{Left: a comb.  The base is bold.  The pixels in the
handle are marked with asterisks, the pixels in the wisdom tooth are
marked with bullets. Right: Voronoi cells. The sides of $\partial P$
are numbered $1\ldots 28$ counterclockwise. The pixels in the
Voronoi cell of a side are marked with the corresponding number.
Voronoi cell of side 3 is a comb; Voronoi cells of sides 6, 11, 17,
25, 28 are empty; cells of sides 1, 7, 10, 18, 24 are lines,
comprised of just one pixel; cells of the other edges are lines with
more than one pixel.}\figlabel{comb_and_vor}
\end{figure}

An analysis of the structure of the Voronoi partition
under our tie breaking rules gives:

\begin{lemma}\label{lem_forest}
  For a side $e\in\partial P$, the Voronoi cell of $e$ is either a
  line (whose dual graph is a straight path), or a comb, or a
  double-sided comb. By our tie-breaking rule, the combs may appear
  only as the Voronoi cells of horizontal edges. The double-sided
  combs may appear only as the Voronoi cells of (horizontal) edges of
  length~1.
\end{lemma}

Let $p$ be a boundary pixel of $P$, let $e\in\partial P$ be the side
of $p$ such that $p\in\Voronoi e$.  We denote $\vor e$ by
$\mathcal{T}(p)$ or $\mathcal{T}(e)$, indicating that it is a unique
tree (a line or a comb) that has $p$ as the root and $e$ as the base.

\mypara{Lower Bounds.}
We exhibit two lower bounds on the cost of an optimal tour,
one based on the number of pixels and the other
on the Voronoi decomposition of the domain.
At any time let $\snow(R)$ be the set of pixels of $R$ covered with
snow and also, abusing notation, the amount of snow on these pixels.
Let $\distance(R)=\frac{1}{D}\sum_{p\in \snow(R)} \vdist{p}$\,.

\begin{lemma}
  Let $R$ be a subset of $P$ with the snowblower starting from a pixel
  outside $R$.  Then $\snow(R)$ and $\distance(R)$ are lower bounds on
  the cost to clear $R$. \thmlabel{lowerbound}
\end{lemma}

\begin{proof}
  For the snow lower bound, observe that region $R$ cannot be cleared
  with fewer than $\snow(R)$ snowblower moves because each pixel of
  $\snow(R)$ needs to be visited.

  For the distance lower bound, observe that, in order to clear the
  snow initially residing on a pixel $p$, the snowblower has to make
  at least $\vdist{p}$ moves.  When the snow from $p$ is carried to
  the boundary of $P$ and thrown away, the snow from at most $D-1$
  other pixels can be thrown away simultaneously.  Thus, a region $R$
  cannot be cleared with fewer than $\dist(R)$ moves. \qed
\end{proof}

\mypara{NP-Completeness.}%
It is known (\cite{ips-hpgg-82, pv-tgprt-84}) that the Hamiltonian
path problem in cubic grid graphs is NP-complete. The problem can be
straightforwardly reduced to SBP. If $G$ is a cubic grid graph,
construct an (integral orthohedral) domain $P$ such that $G=G_P$.
Since $G_P$ is cubic, each pixel $p\in P$ is a boundary pixel, thus,
the snowblower can throw the snow away from $p$ upon entering it.
Hence, SBP on $P$ is equivalent to TSP on $G$, which has optimum
less than $n+1$ iff $G$ is Hamiltonian (where $n$ is the number of
nodes in $G$).  The reduction works for any $D\geq 1$.

The algorithms proposed in this paper show that any domain can be
cleared using a set of moves of cardinality polynomial in the number of pixels
in the domain, assuming $D\geq 2$. Thus, we obtain

\begin{theorem}
If $D\geq 2$, the SBP is NP-complete, both in the default model and in the adjustable throw model, for inputs that are polygonal domains with holes.
\end{theorem}

\section{Approximation Algorithm for the Default
Model}\label{sec_default}

In this section we give an 8-approximation algorithm for the case
when the snow can be thrown in {\em all four} directions.  We first
show how to clear a line efficiently with the operation called {\em
line-clearing}.  We then introduce another operation, the {\em
brush}, and show how to clear a comb efficiently with a sequence of
line-clearings and brushes.  Finally, we splice the subtours through
each line and comb into a larger tour, clearing the entire domain.
The algorithm for the default model, developed in this section,
serves as a basis for the algorithms in the other models.

\mypara{Clearing a Line.}
Let $\mathcal{L}$ be a line of pixels; let $p$ and $e$ be its root
and the base.  We are interested in clearing lines for which the
base is a boundary side, i.e., $e\in\partial P$.  Let
$\ell=\size(\mathcal{L})$; let the first $J$ pixels of $\mathcal{L}$
counting from $p$ be clear.
We assume that $p$ is already clear ($J>0$); the snow from it was
thrown away through the side $e$ as the snowblower first entered
pixel $p$. Let $\mathcal{L}|J$ denote $\mathcal{L}$ with the $J$
pixels clear;  let $\ell - J=kD+r$.\footnote{For ease of
presentation, we adapt the following convention. For
$d\in\{D,\floor{D/2}\}$ and an integer $w$ we understand the
equality $w=ad+b$ as follows: $b$ and $a$ are the remainder and the
quotient, respectively, of $w$ divided by~$d$.} Denote by $(\mathcal{L}|J)_{D}$ the
first $kD$ pixels of $\mathcal{L}|J$ covered with snow; denote by
$\mathcal{L}_r$ the last $r$ pixels on $\mathcal{L}|J$.  The idea of
decomposing $\mathcal{L}|J$ into $(\mathcal{L}|J)_{D}$ and
$\mathcal{L}_r$ is that the snow from $(\mathcal{L}|J)_{D}$ is
thrown away with $k$ ``fully-loaded'' throws, and the snow from
$\mathcal{L}_r$ is thrown away with (at most one) additional
``under-loaded'' throw.

We clear line $\mathcal{L}$ starting at $p$ by moving all the snow
through the base $e$ and returning back to $p$. The basic clearing
operation is a back throw.  In a back throw 
the snowblower, entering a pixel $u$ from
pixel $v$, throws $u$'s snow backward onto $v$. Starting from $p$,
the snowblower moves along $\mathcal{L}$ away from $p$ until either
the snowblower moves through $D$ pixels covered with snow or the
snowblower reaches the other end of $\mathcal{L}$; this is called
the {\em forward pass}. Next, the snowblower makes a U-turn and
moves back to $p$, pushing all the snow in front of it and over $e$;
this is called the {\em backward pass}. A forward and backward pass
that clears exactly $D$ units of snow is called a {\em $D$-full
pass}.

\begin{lemma}\label{lem_line_apx}
For arbitrary $D\ge4$ the line-clearing cost is at most
$2\snow(\mathcal{L}\setminus p) + 4\dist(\mathcal{L}|J)$. For
$D=2,3$ the line-clearing cost is at most
$2\snow(\mathcal{L}\setminus p) + 2\dist(\mathcal{L}|J)$.
If every
pass is $D$-full, the cost is $4\dist(\mathcal{L}|J)$ for $D\ge4$
and $2\dist(\mathcal{L}|J)$
for $D=2,3$. 
\end{lemma}
\begin{proof}
The clearing cost is
$\cost(\mathcal{L}|J) = \cost((\mathcal{L}|J)_{D}) +
\cost(\mathcal{L}_r)  =\sum_{i=1}^k2(J-1+iD)+ 2(\ell-1) =
2kJ+Dk(k+1)-2k+2(\ell-1)$.
The {\em snow} lower bound of $\mathcal{L}\setminus p$ is
$\snow(\mathcal{L}\setminus p) =\ell-1$.
The {\em distance} lower bound of $(\mathcal{L}|J)_D$ is
$\dist((\mathcal{L}|J)_D) = \frac{1}{D}\sum_{i=1}^{kD}(J+i) =
kJ+{k(kD+1)}/{2}$.

Thus,
$$
\cost(\mathcal{L}|J) = 2\snow(\mathcal{L}\setminus p) +
\left(2+\frac{D-3}{J+(Dk+1)/{2}}\right)\dist((\mathcal{L}|J)_D)
$$
If every pass is a $D$-full pass, then $\cost(\mathcal{L}_r)=0$.
Therefore, $\cost(\mathcal{L}|J) = \cost((\mathcal{L}|J)_D) =
\left(2+\frac{D-3}{J+(Dk+1)/{2}}\right)\dist((\mathcal{L}|J)_D)$.\qed
\end{proof}

\mypara{Clearing a Comb.}
Let $\mathcal{C}$ be a comb with the root $p$, base $e$, and handle
$\mathcal{H}$ of length $H$. Let $\ell_1\ldots \ell_H$ be the
lengths of the teeth of the comb. Since we are interested in
clearing combs for which the base $e$ is a boundary side
($e\in\partial P$), we assume that pixel $p$ is already clear
--- the snow from it was thrown away through $e$ as the snowblower
first
entered~$p$. 

Our clearing strategy is as follows.  While there exists a line
$\mathcal{L}\subset\mathcal{C}$ rooted at $p$, such that
$\snow(\mathcal{L})\ge D$, we perform as many $D$-full passes on
$\mathcal{L}$ as we can. When no such $\cal L$ remains, we call the
comb \emph{brush-ready} and we use another clearing operation, the
\emph{brush}, to finish the clearing.

A brush, essentially, is a ``capacitated'' depth-first-search. Among
the teeth of a brush-ready comb that are not fully cleared, let $t$
be the tooth, furthest from the base.  In a brush, we move the
snowblower from $p$ through the handle, turn into $t$, reach its
tip, U-turn, come back to the handle (pushing the pile of snow),
turn onto the handle, move by the handle back towards $p$ until we
reach the next not fully cleared tooth, turn onto the tooth, and so
on. We continue clearing the teeth one-by-one in this manner until
$D$ units of snow have been moved (or all the snow on the comb has
been moved). Then we push the snow to $p$ through the handle and
across $e$. This tour is called a {\em brush} (\figref{brush}).

\begin{figure*}
\centering
\begin{picture}(400,100)
\multiput(-10,0)(130,0){3}{
\multiput(50,10)(10,0){5}{\multiput(0,0)(0,10){7}{\usebox{\sq}}}
\multiput(10,10)(10,0){4}{\multiput(0,0)(0,10){2}{\usebox{\sq}}}
\multiput(0,30)(10,0){5}{\multiput(0,10)(0,10){2}{\usebox{\sq}}}
\linethickness{2pt}\put(90,10){\line(1,0){10}}\thinlines%
}%
\multiput(1,11)(10,0){3}{$\lightgray\blacksquare$}
\multiput(1,21)(10,0){0}{$\lightgray\blacksquare$}
\multiput(41,31)(10,0){2}{$\lightgray\blacksquare$}
\multiput(-9,41)(10,0){1}{$\lightgray\blacksquare$}
\multiput(-9,51)(10,0){2}{$\lightgray\blacksquare$}
\put(130,0){\multiput(1,11)(10,0){3}{$\lightgray\blacksquare$}
\multiput(1,21)(10,0){0}{$\lightgray\blacksquare$}
\multiput(41,31)(10,0){2}{$\lightgray\blacksquare$}
\multiput(-9,41)(10,0){1}{$\lightgray\blacksquare$}
\multiput(-9,51)(10,0){2}{$\lightgray\blacksquare$}}
\thicklines\put(218,15){\line(0,1){42}}\thinlines
\put(218,32){\vector(0,1){5}}%
\put(218,57){\line(-1,0){90}}\put(128,57){\line(0,-1){3}}
\put(128,54){\line(1,0){84}}
\thicklines\put(212,54){\line(0,-1){6}}\thinlines
\put(212,48){\line(-1,0){84}}\put(128,48){\line(0,-1){5}}
\put(128,43){\line(1,0){84}}
\thicklines\put(212,43){\line(0,-1){6}}\thinlines
\put(212,37){\line(-1,0){24}}\put(188,37){\line(0,-1){5}}
\put(188,32){\line(1,0){24}}
\thicklines\put(212,32){\line(0,-1){17}}\thinlines
\put(212,24){\vector(0,-1){5}}
\put(260,0){\multiput(1,11)(10,0){3}{$\lightgray\blacksquare$}
\multiput(1,21)(10,0){0}{$\lightgray\blacksquare$}
\multiput(41,31)(10,0){1}{$\lightgray\blacksquare$}}
\end{picture}
\mycaption{Left: a brush-ready comb.  The snow is shown in light
gray.  Center: a brush, $D=4$; the part of the brush, traveling
through the handle, is bold. Right: the comb after the brush.}
\figlabel{brush}
\end{figure*}

\begin{lemma}\label{lem_comb_apx}
For arbitrary $D\ge4$ the comb $\mathcal{C}$ can be cleared at a
cost of at most $4\snow(\mathcal{C}\setminus p) +
4\dist(\mathcal{C}\setminus p)$ (at most
$4\snow(\mathcal{C}\setminus p) + 2\dist(\mathcal{C}\setminus p)$
for $D=2,3$).
\end{lemma}
\begin{proof}
If $\snow(\mathcal{C}\setminus p)<D$, then the cost of clearing is
just $2\snow(\mathcal{C}\setminus p)$, so suppose,
$\snow(\mathcal{C}\setminus p)\ge D$. Let $B$ be the number of
brushes used; let $\mathcal{B}$ be the set of pixels cleared by the
brushes.  For $b=1\ldots B$ let $t_b$ and $t_b'$ be the first and
the last tooth visited during the $b${th} brush. For
$b\in\set{1\ldots B-1}$ the $b$th brush enters at least 2 teeth, so
$t_b>t_b'\ge t_{b+1}$.

Each brush can be decomposed into two parts: the part, traveling
through the teeth, and the part, traveling through the handle (see
\figref{brush}). Since each tooth is visited during at most 2
brushes, the length of the first part is at most 4 times the size of
all teeth, that is, $4\size(\mathcal{C}\setminus\mathcal{H})$. The
total length of the second part of all brushes is
$2\sum_{b=1}^B(t_b-1)$. Thus, the cost of the ``brushing'' is
\begin{equation}\label{eq_cost_B}
\cost(\mathcal{B})\le 2\sum_{b=1}^B(t_b-1) +
4\size(\mathcal{C}\setminus\mathcal{H}) \le 2\sum_{b=2}^Bt_b +
4\snow(\mathcal{C}\setminus p)-2
\end{equation}
since $t_1\le H$ and $H\ge2$ (for otherwise $\mathcal{C}$ is a
line).

There are exactly $D$ pixels cleared during each brush
$b\in\set{0\ldots B-1}$, and each of these pixels is at distance
at least $t_{b'}$ from the base of the comb.  Thus, the
\emph{distance} lower bound of the pixels, cleared during brush $b$,
is at least $t_{b'}$. Consequently, the \emph{distance} lower bound
of $\mathcal{B}$
\begin{equation}\label{eq_dist_B}
\dist({\mathcal{B}})\ge\sum_{b=1}^{B}t_{b'} \ge
\sum_{b=1}^{B-1}t_{b+1} = \sum_{b=2}^Bt_b
\end{equation}
From (\ref{eq_cost_B}) and (\ref{eq_dist_B}), $\mathcal{B}$ can be
cleared at a cost of at most $2\dist(\mathcal{B}) +
4\snow(\mathcal{C}\setminus p)$.

Let $\mathcal{P}\subseteq\cal C$ be the pixels, cleared during the
line-clearings.  By our strategy, during each line-clearing, every
pass is $D$-full; thus, by Lemma~\ref{lem_line_apx}, $\mathcal{P}$
can be cleared at a cost of at most $4\dist(\mathcal{P})$ (or
$2\dist(\mathcal{P})$ if $D=2,3$). Since $\mathcal{P}$ and
$\mathcal{B}$ are snow-disjoint and
$\mathcal{P}\cup\mathcal{B}=\mathcal{C}\setminus p$, the lemma
follows.
\qed
\end{proof}

The above analysis is also valid in the case when the handle is
initially clear. This is the case when the second side of a
double-sided comb is being cleared. Thus, a double-sided comb can be
cleared within the same bounds on the cost of clearing.

\mypara{Clearing the Domain.}
Now that we have defined the operations which allow us to clear
efficiently lines and combs, we are ready to present the algorithm
for clearing the domain.
\begin{theorem}\label{thm_apx_back_throw}
For arbitrary $D\ge4$ (resp., $D=2,3$) an 8-approximate (resp.,
6-approximate) tour can be found in polynomial time.
\end{theorem}
\begin{proof}
Let $p_1,\ldots,p_M$ be the boundary pixels of $P$ as they are
encountered when going around the boundary of $P$ counterclockwise
starting from $g=p_1$; let $e_1,\ldots,e_M\in\partial P$ be the
boundary sides of $p_1,\ldots,p_M$ such that $e_i= Ve(p_i)$,
$i=1\ldots M$.  
The polygon $P$ can be decomposed into disjoint trees $\mathcal
T(p_1),\ldots,\mathcal T(p_M)=\mathcal T(e_1),\ldots,\mathcal
T(e_M)$ with the bases $e_1\ldots,e_M$, where each tree
$\mathcal{T}(e_i)$ is either a line or a comb.

Our algorithm clears $P$ tree-by-tree starting with
$\mathcal{T}(e_1)=\mathcal{T}(g)$.  By Lemmas~\ref{lem_line_apx}
and~\ref{lem_comb_apx}, for $i=1\ldots M$, the tree
$\mathcal{T}(p_i)\setminus p_i$ can be cleared at a cost of at
most $4\snow(\mathcal{T}(p_i)\setminus
p_i)+4\dist(\mathcal{T}(p_i)\setminus p_i)$ starting from $p_i$ and
returning to $p_i$.  Since $\bigcup_1^M\mathcal{T}(p_i)\setminus
p_i=P\setminus\{p_1\ldots p_M\}$, the interior of $P$ can be cleared
at a cost of at most $\cost(P\setminus\{p_1\ldots
p_M\})=4\snow(P\setminus\{p_1\ldots p_M\}) +
4\dist(P\setminus\{p_1\ldots p_M\}) \le 4\snow(P\setminus
g)+4\dist(P\setminus g)-4M+4$.
Finally, the tours clearing the interior of $P$ can be spliced into
a tour, clearing $P$ at a cost of at most $2M$.
Since the optimum is at least $\snow(P\setminus g)$ and is at least
$\dist(P\setminus g)$, the theorem follows.\qed
\end{proof}

\section{Other Models}\label{sec_restricted}
In this section we give approximation algorithms for the case when
the throw direction is restricted. Specifically, we first consider
the adjustable throw direction formulation. This is a convenient
case for the snowblower operator who does not want the snow thrown
in his face.  We then consider the fixed throw direction model,
which assumes that the snow is always thrown to the right.

We remark that the low approximation factors of the algorithms for
the default model, presented in the previous section, were due to a
very conservative clearing: the snow from {\em every} pixel $p\in P$
was thrown through the Voronoi side $\vs p$.  Unfortunately, it
seems hard to preserve this appealing property if throwing back is
forbidden. The reason is that the comb in the Voronoi cell
$\Voronoi{e}$ of a boundary side $e\in\partial P$ often has a
``staircase''-shaped boundary (see \figref{stair_case}); clearing
the first ``stair'' in the staircase cannot be done without throwing
the snow onto a pixel of $\Voronoi{e'}$, where $e'\ne e$ is another
boundary side.  That's why the approximation factors of the
algorithms in this section are higher than those in the previous
one.  
\figlater{
\begin{figure}[h!]
\centering
\begin{picture}(100,100)
\thicklines\put(50,0){\line(0,1){20}}
\put(50,20){\line(1,0){20}}\thinlines \put(60,10){$E$}
\multiput(60,20)(-10,0){5}{\usebox\sq}
\multiput(60,30)(-10,0){4}{\usebox\sq}
\multiput(60,40)(-10,0){3}{\usebox\sq}%
{\scriptsize{%
\put(23,23){$p$}\put(23,13){$a$}\put(13,23){$b$}\put(23,33){$c$}
}}
\end{picture}
\mycaption{Part of the comb of $\Voronoi{E}$. $Ve(p)\subseteq E$.
$Ve(a), Ve(b), Ve(c)\nsubseteq E$, so the snow form $a$ must be
thrown onto $\Voronoi{E'}\ne \Voronoi{E}$. Nevertheless, observe
that $\vdist{a}=\vdist{p}-1$ -- one could try to make use of
it.}\figlabel{stair_case}
\end{figure}
} 

\subsection*{Adjustable Throw Direction}
In the adjustable-throw model the snow cannot be thrown backward but
can be thrown in the three other directions.  To give a
constant-factor approximation algorithm for this case, we show how
to emulate line-clearings and brushes avoiding back throws
(\figref{emulate_adj}).  The approximation ratios increase slightly
in comparison with the default model.

\noindent\textbf{Line-clearing.\quad}%
We can emulate a (half of a) pass by a sequence of moves, each with
throwing the snow to the left, forward or to the right
(\figref{emulate_adj}, left and center). Thus, the line-clearing may
be executed in the same way as it was done if the back throws were
allowed.  The only difference is that now the snow is moved to the
base when the snow from only $\lfloor{D}/{2}\rfloor$ pixels (as
opposed to $D$ pixels) of the line is gathered.
\begin{lemma}\label{lem_line_adj}
The line-clearing cost is at most
${3D}/{\lfloor{D}/{2}\rfloor}\dist(\mathcal{L}|J)+
2\snow(\mathcal{L}\setminus p)$.  If every pass is
$\floor{D/2}$-full, the cost is
${3D}/{\lfloor{D}/{2}\rfloor}\dist(\mathcal{L}|J)$.
\end{lemma}
\begin{proof}
Let $\ell - J=k'\floor{D/2}+r'$.  Let
$(\mathcal{L}|J)_{\floor{D/2}}$ be the first
$k'\lfloor{D}/{2}\rfloor$ pixels of $\mathcal{L}|J$, let
$\mathcal{L}_{r'}$ be its last $r'$ pixels.  Then the cost of the
clearing of $\mathcal{L}|J$ is
$\cost(\mathcal{L}|J) =
\cost((\mathcal{L}|J)_{\lfloor{D}/{2}\rfloor}) +
\cost(\mathcal{L}_{r'}) =
\sum_{i=1}^{k'}2(J+i\lfloor{D}/{2}\rfloor)+ 2\ell =
2k'J+\lfloor{D}/{2}\rfloor k'(k'+1)+2\ell$.
The lower bounds are given by $ \snow(\mathcal{L}\setminus p)
=\ell-1 $ and
\begin{equation}\label{eq_dist_line_adj}
\dist((\mathcal{L}|J)_{\lfloor\frac{D}{2}\rfloor})=
\frac{1}{D}\sum_{i=1}^{k'{\lfloor\frac{D}{2}\rfloor}}(J+i)=
\frac{\lfloor\frac{D}{2}\rfloor}{D}\left[k'J+
\frac{k'(k'\lfloor\frac{D}{2}\rfloor+1)}{2}\right]
\end{equation}
Thus,
\begin{equation*}
\cost(\mathcal{L}|J)\le \frac{D}{\lfloor\frac{D}{2}\rfloor}
\left(2+\frac{2+\df k'-k'}{k'J+\frac{\df}{2}k'^2 +
\frac{k'}{2}}\right)\dist(\mathcal{L}\setminus p)
+2\snow(\mathcal{L}\setminus p)\qquad\qed
\end{equation*}
\end{proof}
\begin{figure}
\centering
\begin{picture}(50,50)(30,0)
\multiput(0,0)(0,10){7}{\usebox{\sq}}
\multiput(1,21)(0,10){5}{$\lightgray\blacksquare$} \put(3,13){$s$}
\put(2,15){\line(0,1){50}}\put(2,15){\vector(0,1){25}}
\put(2,65){\line(1,0){6}}%
\put(8,15){\line(0,1){50}}\put(8,65){\vector(0,-1){25}}
\put(13,35){$\rightarrow$}%
\put(30,0){\multiput(0,0)(0,10){7}{\usebox{\sq}}\put(3,13){$s$}
\put(1,1){$\lightgray\blacksquare$}
\put(0,0){\line(1,1){10}}\put(10,0){\line(-1,1){10}}}
\end{picture}
\begin{picture}(50,50)
\multiput(0,0)(40,0){2}{
\multiput(0,0)(10,0){2}{%
\multiput(0,0)(0,10){7}{\usebox{\sq}}}\put(3,13){$s$}}%
\put(23,35){$\rightarrow$}%
\multiput(1,21)(0,10){5}{$\lightgray\blacksquare$}
\multiput(11,1)(0,10){7}{$\lightgray\blacksquare$}
\put(2,15){\line(0,1){20}}\put(2,15){\vector(0,1){10}}
\put(2,15){\line(0,1){20}}\put(2,35){\line(1,0){16}}
\put(18,35){\line(0,-1){10}}\put(18,25){\line(-1,0){10}}
\put(8,25){\line(0,-1){10}}\put(8,25){\vector(0,-1){5}}%
\put(41,1){$\lightgray\blacksquare$}
\multiput(41,41)(0,10){3}{\multiput(0,0)(10,0){2}{$\lightgray\blacksquare$}}
\multiput(51,1)(0,10){2}{$\lightgray\blacksquare$}
\put(40,0){\put(0,0){\line(1,1){10}}\put(10,0){\line(-1,1){10}}}
\end{picture}
\hspace{2cm}
\begin{picture}(100,100)
\multiput(50,10)(10,0){5}{\multiput(0,0)(0,10){7}{\usebox{\sq}}}
\multiput(10,10)(10,0){4}{\multiput(0,0)(0,10){2}{\usebox{\sq}}}
\multiput(10,30)(10,0){5}{\multiput(0,10)(0,10){2}{\usebox{\sq}}}
\put(0,40){\usebox\sq}
\thicklines\put(90,10){\line(1,0){30}}\put(90,10){\line(0,-1){10}}\thinlines%
\put(122,7){$bdP$}
\put(10,0){\multiput(1,11)(10,0){3}{$\lightgray\blacksquare$}
\multiput(1,21)(10,0){0}{$\lightgray\blacksquare$}
\multiput(41,31)(10,0){2}{$\lightgray\blacksquare$}
\multiput(-9,41)(10,0){1}{$\lightgray\blacksquare$}
\multiput(1,51)(10,0){1}{$\lightgray\blacksquare$}}
\put(98,15){\line(0,1){42}}\put(98,15){\vector(0,1){21}}
\put(98,57){\line(-1,0){90}}\put(8,57){\line(0,-1){13}}
\put(8,44){\line(1,0){84}}\put(92,44){\line(0,-1){6}}
\put(92,38){\line(-1,0){34}}\put(58,38){\line(0,-1){15}}
\put(58,23){\line(1,0){34}}\put(92,23){\line(0,-1){7}}
\put(92,16){\line(-1,0){74}}\put(18,16){\line(0,-1){8}}
\put(18,8){\line(1,0){67}}\put(18,8){\vector(1,0){34}}
\put(85,8){\line(0,1){6}}\put(85,14){\line(1,0){10}}
\end{picture}
\mycaption{Emulating line-clearing and brush. The (possible) snow
locations are in light gray; $s$ is the snowblower. Left: forward
and backward passes in the default model; there are $D$ units of
snow on the checked pixel. Center: the passes emulation; there is
(at most) $2\lfloor{D}/{2}\rfloor$ units of snow on the checked
pixel. Right: the snow to be cleared during a brush is in light
gray; there is $\lfloor{D}/{2}\rfloor$ light gray pixels. Together
with each gray pixel, the snow from at most 1 other pixel is moved
--- thus, the brush is feasible.}\figlabel{emulate_adj}
\end{figure}

\noindent\textbf{Brush.\quad}%
Brush also does not change too much from the default case. The
difference is the same as with the line-clearing: now, instead of
clearing $D$ pixels with a brush, only $\lfloor{D}/{2}\rfloor$
pixels are (considered) cleared (\figref{emulate_adj}, right). The
definition of a brush-ready comb is changed too --- now we require
that there is less than $\lfloor{D}/{2}\rfloor$ pixels covered with
snow on each tooth of such a comb.

\begin{lemma}\label{lem_comb_adj}
  A comb can be cleared at a cost of
  {${3D}/{\lfloor{D}/{2}\rfloor}\dist(\mathcal{C}\setminus p) +
    4\snow(\mathcal{C}\setminus p)$}.
\end{lemma}

\begin{proof}
In comparison with the default model (Lemma~\ref{lem_comb_apx})
several observations are in place.  The number of brushes may go up;
we still denote it by $B$. The cost of the brushes
$1\ldots B-1$ does not change. If the $B$th brush has to enter the first
tooth, there may be 2 more moves needed to return to the root of the
comb (see \figref{emulate_adj}, right); hence, the total cost of the
brushing~(\ref{eq_cost_B}) may go up by 2.  The {\em distance}
lower bound (\ref{eq_dist_B}) goes down by
${D}/{\lfloor{D}/{2}\rfloor}$.  The rest of the proof is identical
to the proof of Lemma~\ref{lem_comb_apx} (with
Lemma~\ref{lem_line_adj} used in place of
Lemma~\ref{lem_line_apx}).\qed
\end{proof}
Observe that in fact the snow can be removed from {\em more than}
$\lfloor{D}/{2}\rfloor$ pixels during a brush; we just ignore it for
now in our analysis.  Note that a double-sided comb can also be
cleared in the described way.

\noindent\textbf{Clearing the Domain.\quad}%
As in the default case (Theorem~\ref{thm_apx_back_throw}),

\begin{theorem}\label{thm_apx_adj}
A $(4+{3D}/{\lfloor{D}/{2}\rfloor})$-approximate tour can be found
in polynomial time.
\end{theorem}

\subparagraph{Comment on the Parity of $D$.} We remark that if $D$
is even, the cost of the clearing is the same as it would be if the
snowblower were able to move through snow of depth $D+1$ (the slight
increase of ${6}/(D-1)$ in the approximation factor would be due to
the decrease of the {\em distance} lower bound).

\subsection*{Fixed Throw Direction}
In reality, changing the throw direction requires some effort.  In
particular, a snow {\em plower} does not change the direction of
snow displacement at all.  In this section we consider the fixed
throw direction model, i.e., the case of the snowblower which can
only throw the snow to the right.  We exploit the same idea as in
the previous subsection
--- reducing the problem in the fixed throw direction model to
the problem in the default model.  All we need is to show how to
emulate line-clearing and brush.

In what follows we retain the notation from the previous section.
\begin{lemma}\label{lem_line_fixed}
The line-clearing cost is at most
${24D}/{\lfloor{D}/{2}\rfloor}\dist(\mathcal{L}|J)+
25\snow(\mathcal{L}\setminus p)$.  If every pass is
$\floor{D/2}$-full, the cost is
${24D}/{\lfloor{D}/{2}\rfloor}\dist(\mathcal{L}|J)$.
\end{lemma}
\begin{proof}
We first consider clearing a line whose dual graph is embedded as a
single straight line segment and whose base is perpendicular to the
segment;
we describe the line-clearing,
assuming that the line is vertical.  Next, we extend the solution to
the case when the base is parallel to the edges of the dual graph;
this can only be a horizontal line ---
the first tooth in a \mbox{(double-)}comb.  Finally, we consider
clearing an $L$-shaped line;
this can
only by a tooth together with the (part of the) handle.

\subparagraph{A Line $\cal L$ with $G_{\cal L}\bot e$.} As
in the adjustable-throw case (see \figref{emulate_adj}, left and
center), to clear $\mathcal{L}$ we will need to use the pixels to
the right of $\mathcal{L}$ to throw the snow onto.
Let $p'$ be the pixel, following $p$ in $\partial P$.  Before the
line-clearing is begun, it will be convenient to have $p'$ clear.
Thus, the first thing we do upon entering $\mathcal{L}$ (through
$p$) is clearing $p'$. Together with returning the snowblower to $p$
it takes 2 or 4 moves (\figref{setup_and_inv}, left); we call these
moves {\em the double-base setup}.
\begin{figure}
\centering
\begin{picture}(100,100)
\multiput(0,0)(0,50){2}{%
\multiput(0,20)(10,0){2}{\usebox\sq}
\thicklines\put(0,20){\line(11,0){10}}\thinlines\put(2,12){$e$}
\put(2,23){$s$}
\put(7,28){\line(1,0){11}}\put(7,28){\vector(1,0){10}}
}%
\put(18,78){\line(0,-1){6}}
\put(18,72){\line(-1,0){11}}\put(18,72){\vector(-1,0){6}}
\put(10,10){\usebox\sq}\thicklines\put(10,10){\line(0,1){10}}\thinlines\
\put(18,28){\line(0,-1){16}}\put(18,12){\line(-1,0){5}}
\put(13,12){\line(0,1){11}}\put(13,12){\vector(0,1){6}}
\put(13,23){\line(-1,0){6}}
\end{picture}
\begin{picture}(100,100)
\multiput(0,10)(60,0){2}{\put(2,32){$s$}%
\multiput(0,0)(10,0){2}{%
\multiput(1,41)(0,10){3}{$\lightgray\blacksquare$}
\multiput(0,0)(0,10){7}{\usebox\sq}
}%
}%
\put(70,0){\usebox\sq}
\thicklines%
\put(0,10){\line(1,0){20}}\put(70,10){\line(0,-1){10}\line(-1,0){10}}
\thinlines%
\end{picture}
\mycaption{Left: the double-base setup.  Right:
before the forward pass the snow below the snowblower is cleared on
both lines.} \figlabel{setup_and_inv}
\end{figure}

Then, the following invariant is maintained during line-clearing. If
the snowblower is at a pixel $q\in\mathcal{L}$ before starting the
forward pass, all pixels on $\mathcal{L}$ from $p$ to $q$ are clear,
along with the pixels to the right of them (\figref{setup_and_inv},
right). The invariant holds in the beginning of the line-clearing
and our line-clearing strategy respects it.

Each back throw  is emulated with
5 moves (\figref{backthrow_emul_and_last_push_fixed}, left).  After
moving up by $\floor{D/2}$ pixels (and thus, gathering
$2\lfloor{D}/{2}\rfloor$ units of snow on these $\floor{D/2}$
pixels), the snowblower U-turns and moves towards $p$ ``pushing''
the snow in front of it; a push is emulated with 11 moves
(\figref{fwd_emulate_fixed}).

The above observations already show that the cost of
line-clearing increases only by a multiplicative constant in
comparison with the adjustable-throw case.  A more careful look at
the
\figreftwo{backthrow_emul_and_last_push_fixed}{fwd_emulate_fixed}
reveals that: (1) in the push emulation the first two moves are the
opposites of the last two, thus, all 4 moves may be omitted
-- consequently, a push may be emulated by a sequence of only 7
moves; (2) if the boundary side, following $e$, is vertical, the last
push, throwing the snow away from $P$, may require 9 moves
(\figref{backthrow_emul_and_last_push_fixed}, right); and, (3) when
emulating the last back throw in a forward pass, the last 2 of the 5
moves (the move up and the move to the right in
\figref{backthrow_emul_and_last_push_fixed}, left) can be omitted
-- indeed, during the push emulation, the snowblower may as well
start to the right of the snow (see \figref{fwd_emulate_fixed}).
Thus, a line $\mathcal{L}|J$ can be cleared at a cost of $
\cost(\mathcal{L}|J)  \le 4 \quad +
\sum_{i=1}^{k'}\left(J-1+(i-1)\df+5\df+7(J+i\df-1)\right) +J + 5r' +
7(\ell-1) $.

\begin{figure}
\centering
\begin{picture}(100,60)
\multiput(0,0)(60,0){2}{%
\multiput(0,0)(10,0){2}{%
\put(1,41){$\lightgray\blacksquare$}
\multiput(0,0)(0,10){5}{\usebox\sq}
}%
}%
\multiput(1,31)(10,0){2}{$\lightgray\blacksquare$}
\put(61,21){$\lightgray\blacksquare$}
\put(2,22){$s$}\put(62,32){$s$} \put(35,25){$\rightarrow$}
\put(8,32){\line(1,0){5}}
\put(13,32){\line(0,-1){9}}\put(13,23){\vector(0,1){6}}
\put(13,23){\line(1,0){5}}
\put(18,23){\line(0,1){15}}\put(18,38){\line(-1,0){15}}
\put(3,38){\line(0,-1){10}}\put(3,28){\vector(0,1){8}}%
\put(63,22){\scriptsize2}
\end{picture}
\begin{picture}(50,50)
\multiput(60,0)(60,0){1}{%
\multiput(0,0)(0,10){2}{%
\multiput(0,10)(10,0){2}{\usebox\sq}
}%
\thicklines\put(0,10){\line(11,0){10}}\thinlines
\put(2,2){$e$}\put(1,11){$\lightgray\blacksquare$}
\put(14,18){\line(-1,0){7}}\put(14,18){\vector(-1,0){7}}
\put(7,18){\line(0,1){5}}
\put(7,23){\line(1,0){10}}\put(17,23){\line(0,1){5}}
\put(17,28){\line(-1,0){15}}
\put(2,28){\line(0,-1){13}}\put(2,28){\vector(0,-1){13}}
\put(2,15){\line(1,0){11}}\put(13,12){\line(-1,0){8}}
}%
\put(76,15){\scriptsize$s$}
\put(73,15){\line(1,0){3}}\put(76,15){\line(0,-1){10}}
\put(76,5){\line(-1,0){3}}
\put(73,5){\line(0,1){7}}\put(73,5){\vector(0,1){5}}
\thicklines\put(70,10){\line(0,-1){10}}\thinlines
\end{picture}
\mycaption{Left: emulating back throw. Right: pushing the
$2\floor{D/2}$ units of snow away from $P$ and returning the
snowblower to $p$ may require 9 moves.}
\figlabel{backthrow_emul_and_last_push_fixed}
\end{figure}

\begin{figure}
\centering
\begin{picture}(240,60)(-60,0)
\multiput(-60,0)(60,0){5}{%
\multiput(0,0)(10,0){2}{%
\put(1,41){$\lightgray\blacksquare$}
\multiput(0,0)(0,10){5}{\usebox\sq}
}%
\multiput(1,1)(0,10){2}{$\lightgray\blacksquare$}
\put(2,2){\scriptsize2}
}%
\multiput(-58,12)(60,0){3}{\scriptsize2}
\multiput(-25,25)(60,0){4}{$\rightarrow$}
\multiput(-58,32)(60,0){2}{$s$}
\put(72,32){$s$}\put(132,12){$s$}\put(182,22){$s$}
\multiput(-60,0)(60,0){2}{%
\put(1,21){$\lightgray\blacksquare$}\put(2,22){\scriptsize2}
}%
\put(8,38){\line(1,0){10}}
\put(18,38){\line(0,-1){16}}\put(18,38){\vector(0,-1){6}}
\put(18,22){\line(-1,0){11}}
\put(7,22){\line(0,1){10}}\put(7,32){\line(1,0){7}}
\put(71,21){$\lightgray\blacksquare$}\put(75,22){\scriptsize2}
\put(72,38){\line(-1,0){7}}
\put(65,38){\line(0,-1){15}}\put(65,38){\vector(0,-1){8}}
\put(65,23){\line(1,0){9}}\put(74,23){\line(0,-1){10}}
\put(132,17){\line(0,1){8}}\put(132,17){\vector(0,1){7}}
\put(132,25){\line(-1,0){8}}%
\put(122,12){\scriptsize4}
\put(182,12){\scriptsize4}
\end{picture}
\mycaption{Emulating pushing the snow in front of the snowblower.}
\figlabel{fwd_emulate_fixed}
\end{figure}

\subparagraph{A Line $\cal L$ with $G_{\cal L}||e$.} Consider a
horizontal line, extending to the \emph{left} of the base; such a
line may represent the first tooth of a comb.  The double-base can
be cleared with 8 or 12 moves (see \figref{double_base_hor_left}),
the root can be cleared with 3 moves (see
\figref{root_hor_left_and_db_hor_right}, left) instead of 9 moves
(see \figref{backthrow_emul_and_last_push_fixed}, right); the rest
of the clearing does not change.
\figlater{
\begin{figure}
\centering
\begin{picture}(300,150)
\put(0,100){%
\begin{picture}(300,100)
\put(141,21){$\lightgray\blacksquare$}\put(145,34){$s$}
\multiput(61,21)(0,10){2}{$\lightgray\blacksquare$}
\put(51,31){$\lightgray\blacksquare$}
%
\multiput(0,0)(80,0){3}{
\multiput(20,20)(10,0){5}{%
\multiput(0,0)(0,10){2}{\usebox\sq}
}%
\multiput(21,21)(10,0){3}{%
\multiput(0,0)(0,10){2}{$\lightgray\blacksquare$}
}%
\thicklines%
\put(50,0){\line(0,1){20}}\put(50,20){\line(1,0){20}}
\thinlines%
}
\multiput(52,22)(160,0){2}{$s$}
\multiput(80,5)(80,0){2}{$\rightarrow$}
\multiput(57,23)(80,0){2}{%
\multiput(0,0)(0,5){3}{\line(1,0){9}}
\multiput(9,0)(-9,5){2}{\line(0,1){5}}
}%
%
%
\multiput(57,23)(0,10){2}{\vector(1,0){6}}
\multiput(143,23)(0,10){2}{\vector(-1,0){5}}
\end{picture}
}
\put(0,0){%
\begin{picture}(300,100)

\put(141,21){$\lightgray\blacksquare$}
\multiput(61,21)(0,10){2}{$\lightgray\blacksquare$}
\put(51,31){$\lightgray\blacksquare$}

\multiput(0,0)(80,0){3}{
\multiput(20,20)(10,0){6}{%
\multiput(0,0)(0,10){2}{\usebox\sq}
}%
\multiput(60,10)(10,0){2}{\usebox\sq}
\multiput(21,21)(10,0){3}{%
\multiput(0,0)(0,10){2}{$\lightgray\blacksquare$}
}%
\multiput(71,21)(0,10){2}{$\lightgray\blacksquare$}
\thicklines%
\multiput(50,0)(10,0){2}{\line(0,1){20}} \put(50,20){\line(1,0){10}}
\thinlines%
\put(71,11){$\lightgray\blacksquare$}%
}
\multiput(52,22)(160,0){2}{$s$}
\multiput(84,5)(80,0){2}{$\rightarrow$}
\multiput(57,23)(80,0){2}{%
\put(0,0){\line(1,0){6}}\put(6,0){\line(0,-1){7}}
\put(6,-7){\line(1,0){3}}\put(9,-7){\line(0,1){11}}
\put(9,-7){\line(0,1){11}}
\multiput(9,4)(0,5){2}{\line(-1,0){9}}\put(0,4){\line(0,1){5}}
}%
\multiput(57,23)(0,9){2}{\vector(1,0){6}}
\multiput(143,23)(0,9){2}{\vector(-1,0){5}}
\put(145,34){$s$}
\end{picture}
}
\end{picture}
\caption{Setting up the double-base for clearing a horizontal line
extending to the left of its base. Depending on the direction of the
edge adjacent to the base from the right, there are 8 (above) or 12
(below) moves necessary.}\figlabel{double_base_hor_left}
\end{figure}
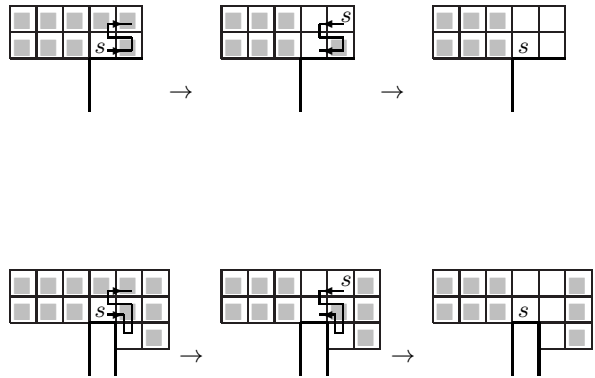

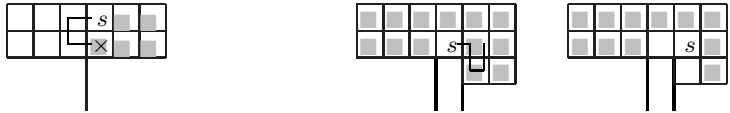
\begin{figure}
\centering
\begin{picture}(100,50)(50,0)
\thicklines%
\put(50,0){\line(0,1){20}}\put(50,20){\line(1,0){10}}
\thinlines%
\multiput(20,20)(10,0){6}{%
\multiput(0,0)(0,10){2}{\usebox\sq}
}%
\multiput(60,20)(0,10){2}{%
\multiput(0,0)(10,0){2}{$\lightgray\blacksquare$}
}%
\put(54,32){$s$}\put(51,21){$\lightgray\blacksquare$}
\put(52,22){$\times$}%
\multiput(52,35)(0,-10){2}{\line(-1,0){9}}
\put(43,35){\line(0,-1){10}}
\end{picture}
\begin{picture}(100,100)(20,0)
\multiput(0,0)(80,0){2}{
\thicklines%
\multiput(50,0)(10,0){2}{\line(0,1){20}}\put(50,20){\line(1,0){10}}
\thinlines%
\multiput(20,20)(10,0){6}{%
\multiput(0,0)(0,10){2}{\usebox\sq}
}%
\multiput(71,11)(0,10){3}{$\lightgray\blacksquare$}
\multiput(20,20)(0,10){2}{%
\multiput(1,1)(10,0){3}{$\lightgray\blacksquare$}
}%
\put(61,31){$\lightgray\blacksquare$}
\multiput(60,10)(10,0){2}{\usebox\sq}
\put(51,31){$\lightgray\blacksquare$}%
}
\put(54,22){$s$} \multiput(61,11)(0,10){2}{$\lightgray\blacksquare$}
\put(58,25){\line(1,0){5}} \multiput(63,25)(5,0){2}{\line(0,-1){10}}
\put(63,15){\line(1,0){5}}
\put(144,22){$s$}
\end{picture}
\caption{Left: clearing the root of a horizontal, extending to the
left, line with 3 moves. There is $2\floor{{D}/{2}}$ units of snow
on the checked pixel. Right: Setting up the double-base for clearing
a horizontal line extending to the right of its
base.}\figlabel{root_hor_left_and_db_hor_right}
\end{figure}
}
%
%

Consider now a horizontal line extending to the \emph{right} of the
base (see \figref{root_hor_left_and_db_hor_right}, right); such a
line may appear as the first tooth in a double-sided comb. The
double-base for such a line can be cleared with 3 moves (see
\figref{root_hor_left_and_db_hor_right}, right); the rest of the
clearing is the same as for the vertical line.

\subparagraph{$L$-shaped Line.} An $L$-shaped line $\mathcal{L}$
consists of a vertical and a horizontal segment.  Each of the
segments can be cleared as described above.

Thus, \emph{any} line $\mathcal{L}|J$ can be cleared at a cost of
at most $ \cost(\mathcal{L}|J)  \le 12 +
\sum_{i=1}^{k'}\left(J-1+(i-1)\df+5\df+7(J+i\df-1)\right) +J + 5r' +
7(\ell-1) $.

Since the \emph{snow} and \emph{distance}
(\ref{eq_dist_line_adj}) lower bounds do not change, the lemma
follows.\qed
\end{proof}

\begin{lemma}\label{lem_comb_fixed}
A comb can be cleared at a cost of
\small{$34\snow(\mathcal{C}\setminus p) +
{24D}/{\floor{D/2}}\dist(\mathcal C\setminus p)$}.
\end{lemma}
\begin{proof}
Brush in the fixed throw direction model can be described easily
using analogy with: a) brush in default and adjustable-throw models
and b) line-clearing in fixed-throw model. As in the
adjustable-throw model, we prepare to clear $\floor{D/2}$ pixels
during each brush. Same as with line-clearing, we setup the
double-base for the comb with at most 12 moves; also, 9 moves per
brush may be needed to push the snow away from $P$ through the base.
Back throw and push can be emulated with 5 and 7 moves
(\figref{backthrow_emul_and_last_push_fixed}, left and
\figref{fwd_emulate_fixed}).  Thus, if the cost of a brush
(\ref{eq_cost_B}) in the default model was, say, $c$, the cost of
the brush in the fixed-throw model is at most $7c+9$. Since any
brush starts with the double-base setup, $c\ge6$; this, in turn,
implies $7c+9\le(51/6)c$. Hence, the cost of clearing $\mathcal{B}$
increases by at most a factor of~${51}/{6}$.

By Lemma~\ref{lem_line_fixed}, the cost of clearing $\cal P$,
$\cost(\mathcal{P}) \le
{24D}/{\floor{D/2}}\dist(\mathcal P)$. 
The \emph{snow} and \emph{distance} lower bounds do not change in
comparison with the adjustable-throw case. The lemma now
follows from simple arithmetic.
\qed
\end{proof}
As in the default and adjustable-throw models (Theorems
\ref{thm_apx_back_throw},\ref{thm_apx_adj}),
\begin{theorem}\label{thm_apx_fixed}
A {\small{$(34+24D/\!\floor{D/2})$}}-approximate tour can be found
in polynomial time.
\end{theorem}

\section{Extensions}\label{sec_extensions}%

\mypara{Polygons with Holes.} Our methods extend to the case in
which $P$ is a polygonal domain with holes. There are two natural
ways that holes may arise in the model.

In the first, the holes represent obstacles for the snowblower
(e.g., walls of buildings that border the driveway).  No snow can be
thrown onto such holes; the holes' boundaries serve as walls for the
motion of the snowblower and for the deposition of snow. Our
algorithm for the default model extends immediately to this
variation. The SBP in restricted-throw models, however, may become
infeasible; it is not always possible for the snowblower to enter a
thin (one-pixel-width) ``channel'' between two walls.

In the second variation, the holes' boundaries are assumed to be the
same ``cliffs'' as the polygon's outer boundary.  It is in fact this
version of the problem that we proved to be NP-complete.  With some
modifications our algorithms work for this variation as well; we
sketch here the necessary modifications.  The boundary, $\partial P$,
of the domain now consists of the boundary pixel sides both on the
{\em outer boundary} of the polygon, denoted $\partial^o\!P$, and on
the boundaries of the holes.  As in the simple polygon case, we build
the Voronoi decomposition of the domain and prepare to clear it
Voronoi-cell-by-Voronoi-cell.  The only problem with it now is that
there is no readily available Hamiltonian cycle through the elements
of $\partial P$. To build a suitable cycle we find a spanning tree,
$\mathbb{T}$, in the graph having a node for each hole and a node for
$\partial^o\!P$ and having an edge between two holes $H_1$ and $H_2$
if there are boundary pixel sides, $e_1\in\partial H_1$ and $e_2\in
\partial H_2$, such that there exist adjacent pixels
$p_1\in\Voronoi{e_1}$ and $p_2\in\Voronoi{e_2}$. Then, after
$\Voronoi{e_1}$ is cleared, we direct the snowblower to start clearing
$\Voronoi{e_2}$.\footnote{The clearing starts from creating a path of
  width 2 from $p_2$ to $e_2$; the path ``bridges'' the holes $H_1$
  and $H_2$.} After that, the snowblower continues to clear the
Voronoi cells of the boundary pixel sides of $H_2$ until another edge
of $\mathbb{T}$, connecting $H_2$ to a hole $H_3$, is encountered (if
$H_2$ is a leaf of $\mathbb{T}$, $H_3=H_1$), and so on.  The cost of
clearing the Voronoi cells of the sides in $\partial P$ does not
change, and the cost of traversing $\mathbb{T}$ is at most twice the size
of $P$; thus, the proposed algorithm remains a constant-factor
approximation algorithm.

\mypara{Nonrectilinear Polygonal Domains.}  Our discussion so far
has assumed that $P$ is integral-orthogonal; in this case, the
snowblower makes axis-parallel movements.  If $P$ is rectilinear,
but not necessarily integral, we proceed as in \cite{afm-aalmm-00}:
first, the boundary of $P$ is traversed once, and then our
algorithms are applied to the remaining part, $P'$, of the domain
$P$.  Every time the snow is thrown away from $P'$, a certain length
(which depends on the throw model) may need to be added to the cost
of the tour; thus, the approximation factors of our algorithms may
increase by an additive constant.


We can also extend our methods to general nonrectilinear domains.
Since the snowblower is not allowed to move outside the domain, care
must be taken about specifying which portion of the domain is actually
clearable. This portion can be found by traversing the boundary of the
domain; then, the accessible portion can be cleared as described
above.

\mypara{Implicit Representation of the Tour.}%
As in \cite{afm-aalmm-00, ahs-oprzp-00}, we make the distinction
between {\em explicit} and {\em implicit} representations of the
domains and snowblower tours.  As mentioned in the introduction, we
assume that a domain is given as the union of pixels; this way the
size of the input to the problem is $O(N)$, where $N$ is the number of
pixels in the domain.  The size of the description of the snowblower
tour produced by our algorithm is polynomial in $N$, i.e., polynomial
in the input size.  Instead, the domain may be given in polygonal
representation, as a list of coordinates of its $n$ vertices; the size
of such a representation is $O(n\log W)$, where $W$ is the largest
coordinate in the input. In principle, $N$ may be $\Omega(W)$, and
hence the length of the description of the snowblower tour may appear
to be exponential in the size of the input. To avoid such a blow-up in
the size of the tour, it is possible to represent the tour succinctly
so that the size of the representation is polynomial in $n\log W$.

Indeed, our algorithms produce tours, comprised of line-clearings and
brushes. Our Voronoi decomposition of the domain is the discretized
version of the Voronoi diagram of the edges of the domain; the latter
is $O(n)$ in size and can be found efficiently (\cite{a-vdsfg-91}).
Given the diagram, it is easy to constrain the (axis-parallel) motion
of the snowblower to stay within a Voronoi cell of an edge: when ``in
doubt'', i.e., when the snowblower is about to enter a pixel,
intersected by an edge of the Voronoi diagram, it can be decided ``in
place'', in constant time (based on the tie breaking rules), whether
entering the pixel will place the snowblower into the Voronoi cell of
another side of~$P$.

The clearing of a Voronoi face in the Voronoi diagram is done
tree-by-tree: first a (double-)comb is cleared (if present), then a
set of lines (if present), then the other comb (if present).  Thus,
finding a short representation of a tour boils down to exhibiting
succinct representations of the tours through a comb and through a set
of lines with adjacent bases comprising a boundary edge of~$P$.  The
descriptions of these tours given in Sections~\ref{sec_default}
and~\ref{sec_restricted} provide such representations.

\mypara{The Vacuum Cleaner Problem.}%
Consider the following problem. The floor -- a polygonal domain,
possibly with holes -- is covered with dust and debris.  The house
is equipped with a central vacuum system, and certain places on the
boundary of the floor (the baseboard) are connected to the ``dustpan
vac'' -- a dust dump location of infinite
capacity~\cite{dustpan-inlet}.  The robotic vacuum cleaner has a
dust/debris capacity $D$ and must be emptied to a dump location
whenever full.  The described problem is equivalent to the SBP in
default throw model, and actually provided the motivation to study
the SBP with throwing backwards allowed.

\mypara{Nonuniform Depth of Snow.}%
Our algorithms generalize straightforwardly to the case in which
some pixels of the domain initially contain more than one unit of
snow. For a problem instance to be feasible it is required that
there is less than $D$ (less than $\floor{D/2}$ in restricted-throw
direction models) units of snow on each pixel. The approximation
ratios in this case depend (linearly) on $D$ (or, in general, on the
ratio of $D$ to the minimum initial depth of snow on~$P$).

\mypara{Capacitated Disposal Region.}%
Another generalization of the problem is to consider the dump
locations to have finite capacities.  If instead of ``cliffs'' at the
boundary of $P$, there is a finite capacity (maximum depth) associated
with each point in the complement of $P$, the SBP more accurately
models some material handling problems, but also becomes considerably
more difficult.  The \emph{snow} lower bound still applies, the
\emph{distance} lower bound transforms to a lower bound based on a
minimum-cost matching between the pixels in $P$ and the pixels in the
complement of $P$.  This problem represents a computational problem
related to ``earth-mover distance'' \cite{cg-emdlbiut-97} and is
beyond the scope of this paper.

\mypara{Open Problems.}%
The complexity of the SBP in simple polygons and the complexity of
the SBP in the fixed-throw model are open.  We also do not have an
algorithm for the case of holes-obstacles and restricted throw
direction; the hardness of this version is also open.

{\small{\baselineskip=12pt
\bibliographystyle{abbrv}
\bibliography{./snowblowing}
}}

\newpage
\section*{Appendix: Figures}
\laterfignow

\printindex
\end{document}